# Electronic and optical properties of La-doped $Sr_3Ir_2O_7$ epitaxial thin-films

M. Souri,[1] J. Terzic,[1,2] J. M. Johnson,[3] J. G. Connell,[1] J. H. Gruenewald,[1] J. Thompson,[1] J. W. Brill,[1] J. Hwang,[3] G. Cao,[2] A. Seo[1]

[1] *Department of Physics and Astronomy, University of Kentucky, Lexington, KY 40506, USA*

[2] *Department of Physics, University of Colorado at Boulder, Boulder, CO 80309, USA*

[3] *Department of Materials Science and Engineering, The Ohio State University, Columbus, OH 43212, USA*

**Abstract**

We have investigated structural, transport, and optical properties of tensile strained $(Sr_{1-x}La_x)_3Ir_2O_7$ ($x$ = 0, 0.025, 0.05) epitaxial thin-films. While High-$T_c$ superconductivity is predicted theoretically in the system, we have observed that all of the samples remain insulating with finite optical gap energies and Mott variable-range hopping characteristics in transport. Cross-sectional scanning transmission electron microscopy indicates that structural defects such as stacking faults appear in this system. The insulating behavior of the La-doped $Sr_3Ir_2O_7$ thin-films is presumably due to disorder-induced localization and ineffective electron-doping of La, which brings to light the intriguing difference between epitaxial thin films and bulk single crystals of the iridates.

I. INTRODUCTION

The Ruddlesden-Popper (RP) phase iridates ($Sr_{n+1}Ir_nO_{3n+1}$ with n = 1, 2, 3, …) have recently received considerable attention due to novel electronic properties [1-7]. One example is the single-layer (n = 1) iridate, $Sr_2IrO_4$, which is theoretically predicted to show a High-$T_c$ superconducting state [8, 9] in the half-filled $J_{eff}$ = 1/2 bands formed by the coexisting strong electron-correlation and spin-orbit interaction [1]. Recent experimental observations of the formation of a *d*-wave gap under surface-electron doping has reinforced this expectation [10, 11]. This observation has encouraged further theoretical and experimental studies utilizing various perturbations such as chemical doping [12-16], pressure [17], and lattice strain [18-20]. The bi-layer (n = 2) iridate, $Sr_3Ir_2O_7$, which is in close proximity to the insulator-to-metal transition with a smaller Mott gap than $Sr_2IrO_4$ [21-28], is another intriguing material that has not been extensively studied under perturbations such as chemical doping and lattice strain. $Sr_3Ir_2O_7$ single crystals show antiferromagnetic ordering ($T_N$ = 285 K [29]) with out-of-plane collinear magnetic moments, in contrast to the in-plane magnetic moments of $Sr_2IrO_4$ and High-$T_c$ cuprates [22, 30]. Recent theoretical studies suggest that a spin flip transition is expected to occur under biaxial strain since the energy difference between the out-of-plane and the in-plane magnetic moments is only a few meV per iridium atom [31]. Hence, investigating electron-doped $Sr_3Ir_2O_7$ thin-films under epitaxial strain could be a promising direction for revealing new electronic states by altering its magnetic structure.

In this paper, we report the structural, transport, and optical properties of La-doped $Sr_3Ir_2O_7$ epitaxial thin-films under coherent tensile-strain. Since the valence state of La is 3+ while that of Sr is 2+, the substitution of La for Sr is widely used for electron doping in many transition metal oxides [32]. However, we have observed that the $Sr_3Ir_2O_7$ epitaxial thin-films with La doping up to about 5 % remain insulating with Mott variable-range hopping characteristics. The La-doped $Sr_3Ir_2O_7$ epitaxial thin-films show overall higher resistivity and slightly larger optical gap energies than $Sr_3Ir_2O_7$. We speculate that La doping might reduce the number of inevitable defects in $Sr_3Ir_2O_7$ epitaxial thin-films such as oxygen vacancies, i.e. two-electron donors. We also have observed structural defects such as stacking faults which commonly appear in RP thin-films through cross-sectional scanning transmission electron microscopy (STEM). Our results imply that, in order to reveal the intrinsic electronic properties of this system, one should look for an effective way of electron doping while reducing chemical and structural disorder.

## II.    METHODS

We have synthesized epitaxial thin-films of $(Sr_{1-x}La_x)_3Ir_2O_7$ ($x$ = 0, 0.025, 0.05) on atomically flat $SrTiO_3$ (STO) (001) substrates [33] by pulsed laser deposition (PLD). The in-plane lattice parameters of the STO substrate and single crystal $Sr_3Ir_2O_7$ are 3.905 Å and 3.896 Å, respectively [21, 34]. Hence, $Sr_3Ir_2O_7$ epitaxial thin-film coherently grown on STO substrate experiences an in-plane tensile strain ($\varepsilon_{xx} = (a_{film} - a_{bulk})/a_{bulk} \times 100$ (%)) of +0.23%. For comparison, $Sr_2IrO_4$ and $SrIrO_3$ thin films, which have pseudocubic in-plane lattice parameters of ~3.89 Å [35] and ~3.96 Å [36], experience in-plane +0.38% tensile [18] and -1.0% compressive [37] strain on a STO substrate, respectively. $Sr_3Ir_2O_7$ (($Sr_{0.95}La_{0.05})_3Ir_2O_7$) thin-films are grown using a ceramic $Sr_3Ir_2O_7$ (($Sr_{0.95}La_{0.05})_3Ir_2O_7$) target with a stoichiometric Sr:Ir ((Sr,La):Ir) ratio of 3:2 (confirmed by energy dispersive x-ray (EDX) spectroscopy) comprised of some additional $IrO_2$ phases (Powder diffraction data not shown.). $(Sr_{0.975}La_{0.025})_3Ir_2O_7$ thin-films are grown by alternating the $Sr_3Ir_2O_7$ and $(Sr_{0.95}La_{0.05})_3Ir_2O_7$ targets. The thin-films are grown with a laser fluence of 1.2 J/cm$^2$ (KrF excimer, $\lambda$ = 248 nm), a substrate temperature of 700 ˚C, and an oxygen partial pressure of 10 mTorr. EDX measurements on the thin-film samples confirms that the average concentration of La ions is consistent with the expected values ($x$ = 0, 0.025, and 0.05). The epitaxial tetragonal structure of our thin-films has been confirmed using four-circle x-ray diffraction (XRD). The transport properties have been measured by using a conventional four-probe method. The in-plane optical absorption spectra of the thin-films have been taken at normal incidence using a Fourier-transform infrared spectrometer and a grating-type spectrophotometer in the photon energy regions of 0.06 – 0.5 eV and 0.5 – 3 eV, respectively.

## III. RESULTS and DISCUSSION

Figure 1 (a) (left panel) shows the $\theta$-$2\theta$ x-ray diffraction (XRD) scans confirming the $c$-axis orientation of $(Sr_{1-x}La_x)_3Ir_2O_7$ thin-films. The enlarged scans in Fig. 1 (a) (right panel) show that the (00$\underline{10}$) reflections of the thin-films are shifted to higher angles as the out-of-plane lattice parameters become smaller by the substitution of $La^{3+}$ on $Sr^{2+}$ sites. The thickness of the thin-films obtained from the XRD interference fringes in the vicinity of the (00$\underline{10}$) peak (Fig. 1 (a) (right panel)) is approximately 25 nm, which is consistent with the thickness obtained from STEM. X-ray reciprocal space mapping (Fig. 1 (b)) shows that the (10$\underline{16}$)-reflection of the thin-films is vertically aligned with the (103)-reflection of the STO substrates indicating that the thin-films are coherently strained to the substrates. The rocking curves are taken from the (10$\underline{16}$)-reflection of the thin-films (Fig 1 (c)) in order to avoid the substrate's truncation rod and its full-width half-maximum (FWHM) is ca. 0.2° for all the thin-films. These large FWHM of thin-film rocking curves compared to the substrate (0.04°) imply that the thin-films have structural disorders such as increased mosaicity. Figure 1 (d) summarizes the out-of-plane lattice parameters ($c$) as a function of La concentration in the $(Sr_{1-x}La_x)_3Ir_2O_7$ thin-films. While the in-plane lattice parameter of the $(Sr_{1-x}La_x)_3Ir_2O_7$ thin-films is constant due to the coherent tensile strain from the STO substrates, La doping decreases the out-of-plane lattice parameter. While $c$-axis contraction with La doping also has been observed in bulk samples of other transition-metal oxides, including iridates, it has generally been associated with increased unit-cell volumes [38-40]. However, note that the ionic radius of $La^{3+}$ (1.03 Å) is smaller than that of $Sr^{2+}$ (1.18 Å) [41]. The increased unit-cell volume with La doping is mostly due to the changes in the valence states of transition metal ions (e.g. from $Ir^{4+}$ to $Ir^{3+}$) and/or the creation of oxygen vacancies. Hence, the decreased $c$-axis lattice parameters of our $Sr_3Ir_2O_7$ thin-films might imply that La-doping does not dope electron carriers effectively in these samples (See also the following discussions about transport properties).

The temperature-dependent resistivity shows that all of the $(Sr_{1-x}La_x)_3Ir_2O_7$ ($x$ = 0, 0.025, 0.05) thin-films exhibit insulating behaviors (Fig. 2 (a)). We have compared the resistivity of these compounds with that of $Sr_2IrO_4$ (purple) and $SrIrO_3$ (orange) thin-films [19, 20, 37], which shows that the $(Sr_{1-x}La_x)_3Ir_2O_7$ thin-films are more (less) insulating than $SrIrO_3$ ($Sr_2IrO_4$) thin-film. The resistivity of the $(Sr_{1-x}La_x)_3Ir_2O_7$ thin-films at room temperature is about $5\times10^{-3}$ Ω·cm, which is almost the same as a previously reported $Sr_3Ir_2O_7$ thin-film [28] but smaller than that of a $Sr_3Ir_2O_7$ single crystal by approximately two orders of magnitude [24]. While the substitution of $La^{3+}$ on $Sr^{2+}$ sites is expected to dope electrons into the system, the resistivity of our thin-films at low temperatures increases by about three orders of magnitude as the La concentration is increased from 0 to 5 %, which is opposite to the metallic behavior of La-doped $Sr_3Ir_2O_7$ single crystals [24, 38]. Hence, our experimental data implies that there are differences in electronic structure between these iridate thin-films and single crystals. To understand the conduction mechanism of these samples, we have considered three transport mechanisms: the thermal activation model ($\rho = \rho_0 e^{\Delta/2k_BT}$, where $\Delta$ is the activation energy and $k_B$ is the Boltzmann constant), the three dimensional (3D) Mott variable-range hopping (VRH) model ($\rho = \rho_0 e^{(T_M/T)^{1/4}}$, where $\rho_0$ is the resistivity coefficient and is $T_M$ the characteristic temperature), and the Efros-Scklovskii (ES) VRH model ($\rho = \rho_0 e^{(T_{ES}/T)^{1/2}}$, where $\rho_0$ is the resistivity coefficient and $T_{ES}$ is the characteristic temperature). Both the thermal activation and ES-VRH model do not fit our experimental data, suggesting that these two models cannot describe the conduction mechanism of this system (model fits not shown). However, as shown in Fig. 2 (b), the 3D Mott-VRH model fits our experimental data very well over a wide temperature range (2 K – 300 K). This implies that the $(Sr_{1-x}La_x)_3Ir_2O_7$ thin-films are a strongly disordered system with localized electrons carriers, similar to $Sr_2IrO_4$ thin-films [42]. Note that our $Sr_3Ir_2O_7$ thin-film is less insulating than single crystal $Sr_3Ir_2O_7$ [24] presumably due to some inevitable oxygen vacancies introduced during the thin-film deposition. However, the La doped $Sr_3Ir_2O_7$ thin-films show larger low-temperature resistivity and $T_M$ values than the $Sr_3Ir_2O_7$ thin-films. The estimated $T_M$ using our resistivity results are 100 K ($Sr_3Ir_2O_7$), 1200 K (($Sr_{0.975}La_{0.025})_3Ir_2O_7$), and 4000 K (($Sr_{0.95}La_{0.05})_3Ir_2O_7$). Since oxygen stoichiometry can be stabilized near La ions [43, 44], we speculate that La doping in our $Sr_3Ir_2O_7$ thin-films might have eliminated some oxygen vacancies reducing electron carriers. Note that a reduction of oxygen vacancies and increased resistivity by La doping has also been reported in some transition-metal oxides such as La-doped $Bi_4Ti_3O_{12}$ single crystals [43, 44].

Optical absorption spectra ($\alpha(\omega)$) confirms the insulating behavior of the $(Sr_{1-x}La_x)_3Ir_2O_7$ thin-films with finite optical gap energies. Figure 3 presents $\alpha(\omega)$ of $(Sr_{1-x}La_x)_3Ir_2O_7$ thin-films together with $Sr_2IrO_4$ [18] and $SrIrO_3$ [37] thin-films. Due to the Reststrahlen band of the STO substrates, ~ 0.2 eV is the lowest photon energy at which optical transmission spectra can be measured. We have performed spectral fits using the minimum set of Lorentz oscillators (black dashed lines), which match well with the experimental spectra. The charge-transfer transitions from O 2$p$ to Ir 5$d$ bands are above 2 eV. Similar to $Sr_3Ir_2O_7$ single crystals [26], $\alpha(\omega)$ of $(Sr_{1-x}La_x)_3Ir_2O_7$ thin-films show a two-peak feature (indicated by $\alpha$ and $\beta$) at low energies, due to a Fano-like coupling between the spin-orbit exciton and inter-site $d$-$d$ transitions within the $J_{eff} = 1/2$ band [16]. In order to extract the optical gap energy, we fit the absorption edge of the $Sr_3Ir_2O_7$ and $(Sr_{0.975}La_{0.025})_3Ir_2O_7$ thin-films using the Wood-Tauc method [45] (Fig. 3 (inset)). In this method, the absorption spectra in the $\alpha > 10^4$ cm$^{-1}$ region is described by the equation below:

$$\alpha \propto \frac{(\omega - E_g)^{\gamma}}{\omega} \qquad (1)$$

where $E_g$ is the optical gap energy and $\omega$ is the photon energy. The estimated optical gaps are about 60 meV and 80 meV for $Sr_3Ir_2O_7$ and $(Sr_{0.975}La_{0.025})_3Ir_2O_7$, respectively. However, the short tail of the $(Sr_{0.95}La_{0.05})_3Ir_2O_7$ spectra at low energy makes it difficult to fit using Wood's method. From our best fit, we obtain $\gamma = 1.5$ for both $Sr_3Ir_2O_7$ and $(Sr_{0.975}La_{0.025})_3Ir_2O_7$, which suggests a direct band gap for the thin-films, in contrast to the indirect band gap ($\gamma = 3.0$) of $Sr_3Ir_2O_7$ single crystals [26] and $Sr_2IrO_4$ thin films [20]. The increase in the optical gap with increasing La concentration is consistent with the systematic increase in the resistivity data of this system.

Cross-sectional STEM of our samples shows structural defects such as stacking faults, which is consistent with Mott-VRH with strong disorder. Figure 4 (a) shows the STEM image of the $Sr_3Ir_2O_7$ thin-film in which the electron beam is incident along the [100]-direction of the (001) STO substrate. There are three noteworthy regions in the STEM data: 1) The ideal bilayer $Sr_3Ir_2O_7$ region with no intergrowth (Fig. 4 (b) and the orange dashed rectangle in Fig. 4 (a)), which is the most widely observed region in large-scale STEM images. 2) the region with a misaligned single SrO layer, i.e. an example of stacking faults, in the $Sr_3Ir_2O_7$ structure (as shown by the blue arrows in Fig. 4 (c) and the yellow dashed rectangle in Fig. 4 (a)). This small region may be structurally similar to $Sr_2IrO_4$. However, since XRD did not show any sign of $Sr_2IrO_4$ phases, the region must be due to structural heterogeneity or defects localized at the nanoscale, and its volume fraction should be negligible. And finally, 3) Figure 4 (d) shows an example of three separate atomic layers that are stacked along the *b*-axis direction (into the page) with vertically shifted SrO layers. The overlap of these three layers is consistent with the region shown by the blue dashed rectangle in the STEM image in Fig. 4 (a). Even though x-ray diffraction shows a single phase epitaxial thin-film, STEM images clearly indicate that there are various misaligned $Sr_3Ir_2O_7$ layers, which affect the transport properties of this system. It is well known that deposition of RP phase thin-films such as $Sr_2RuO_4$ and $Sr_3Ru_2O_7$ can lead to the formation of unit cell fractions in the crystal [46-49]. The formation of these defects causes translation boundary defects in the compound which result in an out-of-plane shift between regions compared with the ideal crystal structure. Although these extra layered intergrowths are not frequently observed in single crystals, they are common in epitaxial thin-films due to the thermodynamic non-equilibrium process of synthesis [49-51].

## IV.    CONCLUSION

We have investigated epitaxial thin films of tensile-strained, La-doped $Sr_3Ir_2O_7$, which are in close proximity to the metal-insulator transition. From transport and optical spectroscopic measurements, the thin-film series shows insulating properties with 3D Mott-VRH, where the resistivity increases by increasing the La concentration. The insulating behavior of the La-doped $Sr_3Ir_2O_7$ thin-films is presumably due to localization of carriers through ineffective electron doping of La and structural defects that are observed in STEM. Note that such structural defects strongly affect materials properties in general. For example, stacking faults can alter or even eliminate superconductivity in $YBa_2Cu_3O_{7-\delta}$ and $Sr_2RuO_4$ [48, 51-54]. Attempts to synthesize superconducting $Sr_2RuO_4$ thin-films have had mixed results [53, 55-58]. The existence of structural defects in this compound limits the in-plane resistivity and quenches the superconductivity [48, 51, 53]. The transport properties of La-doped $Sr_3Ir_2O_7$ thin-films may have been similarly deteriorated due to the observed stacking faults and disorder-induced localization. Therefore, our results highlight a critical role of characterizing and eliminating these defects in epitaxial thin-films to reveal the intrinsic physical properties of the iridates.

## V. ACKNOWLEDGMENTS

We acknowledge the support of National Science Foundation grants DMR-1454200 (for thin-film synthesis and characterizations), DMR-1265162 and DMR-1712101 (for target synthesis), and DMR-1262261 (for infrared spectroscopy).

**FIGURE CAPTIONS**

**FIG. 1.** (a) X-ray $\theta$-$2\theta$ scans of the epitaxial $(Sr_{1-x}La_x)_3Ir_2O_7$ ($x$ = 0, 0.025, 0.05) thin-films grown on STO substrates, where only the (00$l$)-diffraction peaks of the thin-films ($l$ = 4, 6, 10, 12, 14) are visible. The enlarged scans near (00$\underline{10}$) reflections of the thin-films and the (002) reflections of the substrates are shown on the right. The peaks from the substrates are labeled with asterisk (∗) symbols. (b) Reciprocal space map around the (103) reflection of the STO substrates with the (10$\underline{16}$) reflection of $(Sr_{1-x}La_x)_3Ir_2O_7$ ($x$ = 0, 0.025, 0.05) thin-films. (c) The rocking curve scan of $(Sr_{1-x}La_x)_3Ir_2O_7$ thin-films (00$\underline{10}$)-reflection which have a full-width half-maximum of 0.2° for all thin-films. (d) The out-of-plane lattice parameters of the $(Sr_{1-x}La_x)_3Ir_2O_7$ thin-films obtained from the x-ray diffraction are shown as a function of La concentration. The open circle, square, and diamond indicate the out-of-plane lattice parameters of $Sr_3Ir_2O_7$ single crystals from Refs. [21, 29, 34], respectively.

**FIG. 2.** (a) Temperature dependent resistivity of the $(Sr_{1-x}La_x)_3Ir_2O_7$ thin-films, which indicates that all of the thin-films are insulators. The resistivity of $Sr_2IrO_4$ thin-film [20] and a $SrIrO_3$ thin-film [37] on STO substrates is shown for comparison. (b) Logarithmic resistivities versus $T^{-1/4}$, which are offset along the y-axis for clarity, show that all of the $(Sr_{1-x}La_x)_3Ir_2O_7$ thin-films display a 3D Mott-VRH conduction mechanism. The dashed lines indicate the linear fit.

**FIG. 3.** Optical absorption spectra ($\alpha\,(\omega)$) of $(Sr_{1-x}La_x)_3Ir_2O_7$, $Sr_2IrO_4$ (Ref. [20]), and $SrIrO_3$ (Ref. [37]) thin-films at room temperature. The plots are shifted vertically by $10^5$ cm$^{-1}$ for clarity. The dotted black curves are the fit curves using Lorentz oscillators, which match well with the experimental spectra. The inset shows the fitted absorption spectra at low energy using Wood-Tauc's method.

**FIG. 4.** (a) A cross-sectional Z-contrast STEM image of $Sr_3Ir_2O_7$. The brightest spots are Ir atoms; Sr, Ti, and O atoms are faint due to their small atomic (Z) number. (b) The orange dashed rectangular region of the STEM image in (a) with ideal $Sr_3Ir_2O_7$ unit cell. The schematic diagram of the unit cell is shown for clarity. Red dots represent Ir atoms in this and following images. (c) The yellow dashed rectangular region of the STEM image and schematic diagram showing the intergrowth of one single layer of SrO, shown by blue arrows, which causes a structural change from $Sr_3Ir_2O_7$ to $Sr_2IrO_4$. (d) The blue dashed rectangular region of the STEM image and schematic diagram that indicates three separate atomic layers sequentially stacked into the page. The foremost layer (red) is the ideal structure, while the second (blue) and third (green) layers contain vertical intergrowth of single SrO layers as shown by the red arrows. These layers show a non-zero overlap as indicated in the rightmost diagram.

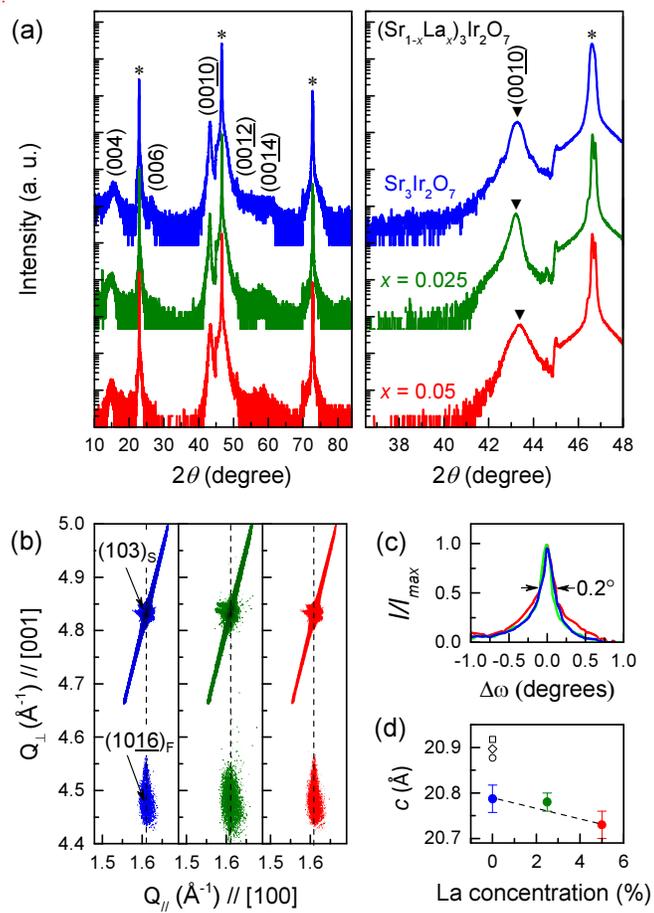

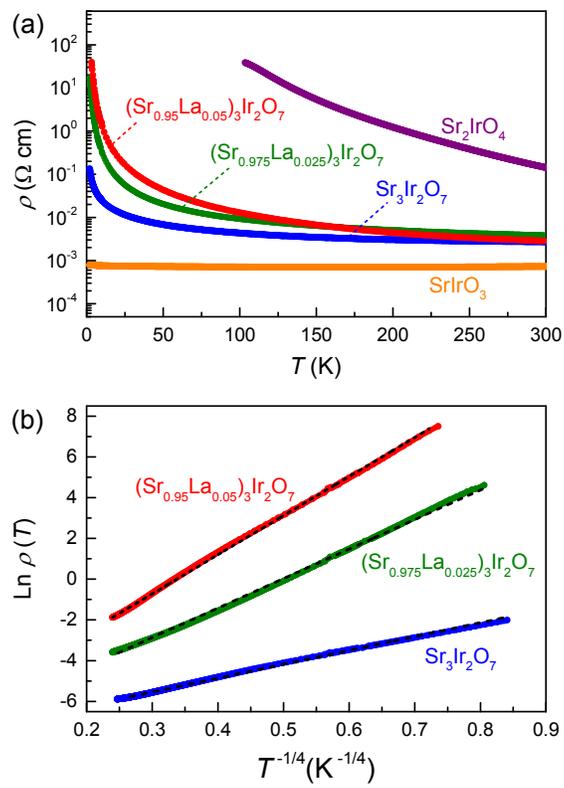

Fig. 2
M. Souri *et al.*

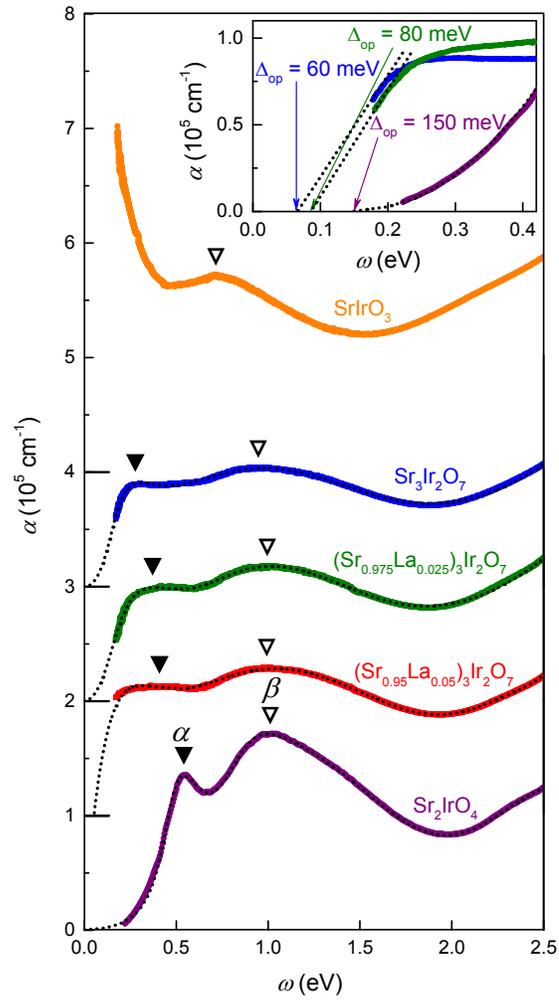

Fig. 3
M. Souri *et al.*

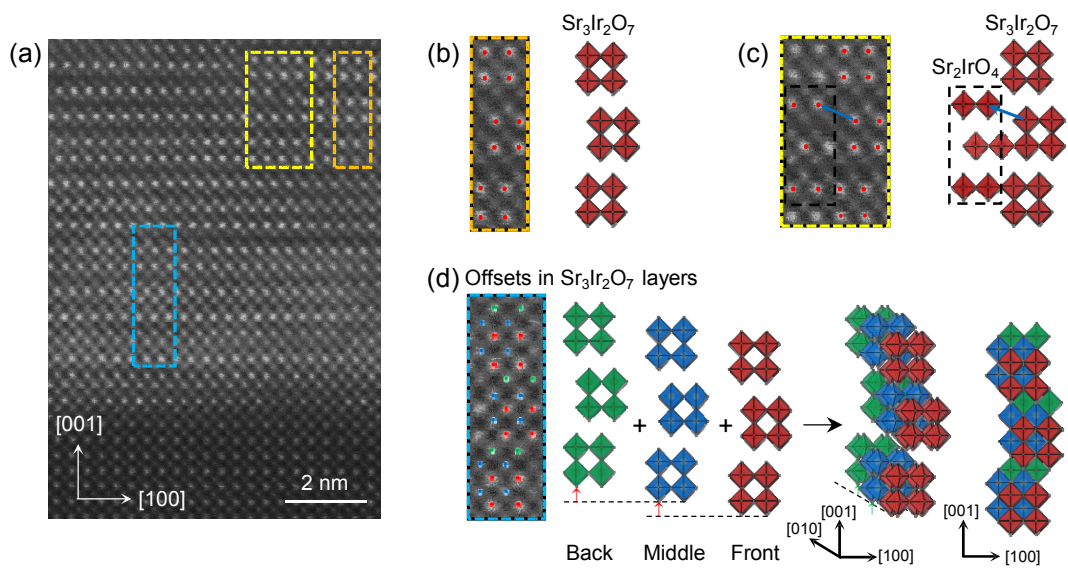

Fig. 4
M. Souri *et al.*